\documentstyle[aps,prabib,twocolumn,epsfig]{revtex}
\begin{document}
\draft
\title{
Spin Environment Engineering for a Persistent Current Qubit 
}
\author{Jacek Dziarmaga$^{1,2}$
        \thanks{e-mail address: dziarmaga@t6-serv.lanl.gov}
}
\address{ 1) Los Alamos National Laboratory, Theory Division T-6, 
             MS B288, Los Alamos, NM87545, USA\\
          2) Institut Fizyki Uniwersytetu Jagiello\'nskiego,
             Krak\'ow, Poland\\
}
\date{January 30, 2001}
\maketitle
\tighten
  
\begin{abstract}

A persistent current qubit has two quantum states with opposite currents
flowing in a superconducting loop. Their magnetic field couple to nuclear
spins. The qubit state is not only perturbed by the spins but it also gets
entangled with the spins' state on a very short timescale. However, when
the same spins are exposed to a strong but less than critical external
magnetic field, then the qubit field is just a small perturbation on top
of the external field and the entanglement with each spin is negligible.
For a qubit which is more microscopic than certain threshold this partial
entanglement results in negligible partial decoherence.

\end{abstract}

\section{Introduction}

  The idea to perform computation with the help of quantum mechanics dates
back to Feynman \cite{feynman}. Research in this direction got a
considerable acceleration after the realisation \cite{NP} that quantum
computers could solve in a polynomial time certain problems which require
nonpolynomial time on classical computers. Two main obstacles to a
universal quantum computer are decoherence and scalability.  Decoherence,
if sufficiently weak, can be dealt with by quantum error correcting codes,
see Ref.\cite{NMR} for an experimental implementation. Scalability
requires that a given technology can be upscaled to a quantum coherent
circuit with hundreds or thousands of quantum gates, see Ref.\cite{review}
for reviews on various proposals of scalable implementations. At the
present stage solid state implementations have most to offer from the
point of view of scalability but at the same time they are most likely to
suffer from decoherence.  The aim of this note is to reduce decoherence in
the persistent current qubit (PCQ) proposal \cite{PCQ}.

  Two remarkable experiments \cite{exp} took place last year where a
superconducting loop broken by Josephson junctions was forced into states
which are quantum superpositions of two states with opposite macroscopic
persistent currents: $|\uparrow\rangle$ and $|\downarrow\rangle$. The two
considered combinations
$|\pm\rangle=|\uparrow\rangle\pm|\downarrow\rangle$ were eigenstates
separated by a gap of $0.1 K$ so it is not surprising that each of them
does not suffer from decoherence at the temperature of $50mK$. A more
general superposition $\alpha|+\rangle+\beta|-\rangle$ would quickly
decohere into a mixed state $|\alpha|^2 |+\rangle\langle+| + |\beta|^2
|-\rangle\langle-|$. To use a persistent current qubit for quantum
computation any superposition must be free from decoherence. A design of a
more microscopic scalable PCQ was proposed in Ref.\cite{PCQ}.  These
authors went on to estimate decoherence times from different sources
relevant for their design \cite{PCQdec} and found out that the most
dangerous are nuclear spins which couple to the magnetic field induced by
persistent currents. Similar conclusion is reached in Ref.\cite{stamp}. In
the calculations of Ref.\cite{PCQdec} the spins are treated as random
static background fields as may be justified by the long relaxation time
of nuclear spins which is of the order of minutes. Such a static
background can be measured at the beginning of the quantum computation and
compensated for by an adjustable counterterm in the qubit Hamiltonian. In
this note we re-address this problem and find out that spins cannot be
treated as a mere random background because the qubit state quickly gets
entangled with the spins state. This observation is in agreement with
Ref.\cite{stamp}. What is more, because of the large number of involved
spins, even for the long relaxation time the spins background performs a
random walk which leads to dephasing in the qubit state. The former more
acute entanglement problem can be treated by exposing the nuclear spins to
a strong external magnetic field $B$ parallel to the plane of the
superconducting loop and penetrating through its thin layer, see Fig.1.

\begin{figure}\label{design}
\centerline{\epsfxsize=6 cm \epsfbox{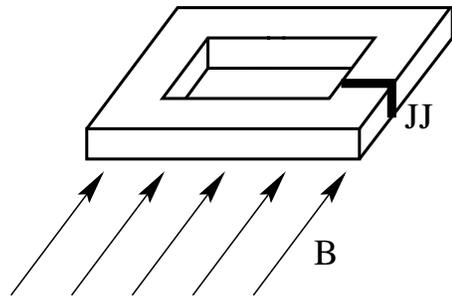}}
\caption{
The idea of how the external magnetic field $B$ can be introduced
into a superconducting loop. Only one Josephson junction (JJ)
is marked.
}
\end{figure}
$B$ has to be less than the critical field in a type I superconducting
film. If the $N$ nuclear spins are exposed to a magnetic field $b$ from
the persistent current which is much weaker than $B$, as can be
characterized by a quality factor

\begin{equation}
Q=\frac{B^2}{Nb^2}\;, 
\end{equation}
then the entanglement between the qubit state and the spins state is
negligible and leads to negligible {\it partial decoherence}. The means
which treat entanglement incidentally also reduce the latter random walk
problem, which can be furter eliminated by increasing the spin relaxation
time $T_r$ which depends exponentially on temperature.

\section{ Single spin environment }

  We begin with an elementary example where the qubit is coupled to just
one environmental $1/2$ spin. For the purpose of this paper the spin-1/2
is a sufficient representation of the nuclear spin-9/2 of niobium. The
spin is coupled to the external magnetic field $B$; its Hamiltonian is

\begin{equation}
H_{\rm{E}} = B \sigma_z^{\rm{E}} \;.
\end{equation}
The current of the qubit induces a magnetic field which at the location of
the environmental spin has strength $b$. For the time being we assume that
the field is perpendicular to the external field in, say, $x$-direction.
The Hamiltonian of interaction between the qubit and the spin is

\begin{equation}
H_{\rm{Q-E}} = b \sigma_{z} \sigma_{x}^{\rm{E}} \;.
\end{equation}
The external magnetic field $B$, which is in plane of the 
superconducting loop, does not couple to the qubit state. 
For the sake of simplicity we set the qubit Hamiltonian
$H_{\rm{Q}}=0$, what does not change the general conclusion
of this note. The initial state of $Q$ and $E$ is a product

\begin{equation}\label{psi0}
|\psi(0)\rangle\left( \alpha |\uparrow\rangle + 
       \beta  |\downarrow\rangle \right)
|\downarrow_{\rm{E}}\rangle\;.
\end{equation}
The states $|\downarrow_{(\rm{E})}\rangle,|\uparrow_{(\rm{E})}\rangle$ are
$\pm 1$ eigenstates of $\sigma_z^{(\rm{E})}$. Under the total Hamiltonian
$H_{\rm{Q-E}}+H_{\rm{E}}$ the initial $|\psi(0)\rangle$ evolves into

\begin{eqnarray}\label{psit}
|\psi(t)\rangle&=&
\alpha
|\uparrow\rangle 
e^{-it( +b \sigma_x^{\rm{E}} + B \sigma_z^{\rm{E}} )}
|\downarrow_{\rm{E}}\rangle + \nonumber\\
&& 
\beta
|\downarrow\rangle 
e^{-it( -b \sigma_x^{\rm{E}} + B \sigma_z^{\rm{E}} )}
|\downarrow_{\rm{E}}\rangle \;.
\end{eqnarray}
At the same time the initial pure reduced density matrix of 
the qubit

\begin{equation}\label{rho0}
\rho(0) = 
\pmatrix{
\alpha^{\star}\alpha  &  \alpha\beta^{\star} \cr
\alpha^{\star}\beta   &  \beta^{\star}\beta  \cr} 
\end{equation}
evolves into 

\begin{equation}\label{rhot}
\rho(t) = 
\pmatrix{
    \alpha^{\star}\alpha  &  O(t)\alpha\beta^{\star} \cr
O(t)\alpha^{\star}\beta   &      \beta^{\star}\beta  \cr}\;, 
\end{equation}
where the overlap $O(t)$ is given by

\begin{eqnarray}
O(t) & \equiv &
\langle\downarrow_{\rm{E}}|
e^{+it( +b \sigma_x^{\rm{E}} + B \sigma_z^{\rm{E}} )}
e^{-it( -b \sigma_x^{\rm{E}} + B \sigma_z^{\rm{E}} )}
|\downarrow_{\rm{E}}\rangle = \nonumber\\
&& 1-\frac{2b^2}{\Omega^2}\sin^2\sqrt\Omega t \;. 
\end{eqnarray}
Here $\Omega=\sqrt{B^2+b^2}$. $\rho(t)$ is mixed when $|O(t)|<1$.  The
entanglement with a single spin is reversible because there are periodic
revivals at the times when $O(t)=1$. We note that for $b^2 \ll B^2$ the
overlap $O(t)$ remains $1$ and the qubit state stays pure. 

  Even just one environmental spin can lead to irreversible full
decoherence when the spin interacts with its thermal bath. In the absence
of the qubit, thermalisation of the spin density matrix $\rho_{\rm{E}}$
towards $\frac12 I$ can be described by a master equation

\begin{equation}
T_r\frac{d}{dt}\rho_{\rm{E}}=
\sigma_- \rho_{\rm{E}} \sigma_+ +
\sigma_+ \rho_{\rm{E}} \sigma_- -
\rho_{\rm{E}} \;,
\end{equation}
where $\sigma_{\mp}=(\sigma_x^{\rm{E}}\mp i\sigma_y^{\rm{E}})/2$ are
spin-$1/2$ lowering/raising operators and $T_r$ is the spin relaxation
time of the order of minutes. This master equation can be formally
unravelled by a stochastic evolution of the spin state. The spin state
makes a down(up) flip within time $dt$ with a probability given by
$dt/T_r$ times the probability that the spin is in the
$|\uparrow_{\rm{E}}\rangle(|\uparrow_{\rm{E}}\rangle)$ state.  The master
equation describes evolution of an average over such stochastic
trajectories of the spin state. 

  Let us apply this stochastic unravelling to the state of the spin and
the qubit. The $|\psi(t)\rangle$ in Eq.(\ref{psit}) can be rewritten as

\begin{eqnarray}\label{psitt}
&& 
|\psi(t)\rangle=             
\left( \frac{-ib}{\Omega}
       \sin\Omega t      \right)
\left( \alpha |\uparrow\rangle -
       \beta  |\downarrow\rangle \right)
|\uparrow_{\rm{E}}\rangle + 
\nonumber\\ 
&&
\left( \cos\sqrt\Omega t+
       \frac{iB}{\Omega}\sin\Omega t      \right)
\left( \alpha |\uparrow\rangle+
       \beta  |\downarrow\rangle \right)
|\downarrow_{\rm{E}}\rangle
\end{eqnarray}
The up transition at the time $t$ can be described as a projection of
$|\psi(t)\rangle$ on $\langle\downarrow_{\rm{E}}|$ followed by a spin flip
\\
$|\downarrow_{\rm{E}}\rangle\rightarrow|\uparrow_{\rm{E}}\rangle$.  As a
result $|\psi(t)\rangle$ jumps to

\begin{equation}
|\psi(t^+)\rangle= 
\left( \alpha |\uparrow\rangle +
       \beta  |\downarrow\rangle  \right)
|\uparrow_{\rm{E}}\rangle  \;.
\end{equation}
The resulting state of the system is the same as its initial state: this
up transition causes no decoherence. 
  
   The probability of the down transition in the interval $(t,t+dt)$ is
$dt/T_r$ times the probability that $|\psi(t)\rangle$ is in the
$|\uparrow_{\rm{E}}\rangle$ state which is
$|\langle\uparrow_{\rm{E}}|\psi(t)\rangle|^2
 \approx (b^2/B^2)\sin^2 Bt \approx b^2/2B^2$, where we assume 
$b^2\ll B^2$. The probability is $dtb^2/2B^2T_r$. The down transition
is a projection on $\langle\uparrow_{\rm{E}}|$ followed by a spin flip 
$|\uparrow_{\rm{E}}\rangle\rightarrow|\downarrow_{\rm{E}}\rangle$.
$|\psi(t)\rangle$ jumps to

\begin{equation}
|\psi(t^+)\rangle=
\left( \alpha |\uparrow\rangle_{\rm{S}}-
       \beta  |\downarrow\rangle_{\rm{S}}  \right)
|\downarrow\rangle_{\rm{E}} \;. 
\end{equation}
The resulting system state has a flipped sign of $\beta$ as compared to
the initial system state (\ref{psi0}). This sign flip also flips the sign
of the off-diagonal elements in the reduced qubit density matrix, compare
Eqs.(\ref{rho0},\ref{rhot}). The evolution of $\rho(t)$ including spin
relaxation is obtained after an average over stochastic trajectories. The
off-diagonal elements of $\rho(t)$ average to zero on a timescale of
$\tau^{(2)}_B=2T_rB^2/b^2$. This is the time of full decoherence due to
spin relaxation. We note that for $B^2\gg b^2$ the time is much longer
than $T_r$.

\section{ N-spin environment }

  Decoherence becomes more efficient when there are many copies of the
environmental spin numbered by $i=1,\dots,N$.  Interaction Hamiltonians
are $H_{\rm{Q-E}_i}=b_i\sigma_z\sigma_x^{\rm{E}_i}$. The initial product
state

\begin{equation}
|\psi(0)\rangle=
\left( \alpha |\uparrow\rangle + 
       \beta  |\downarrow\rangle \right)
\prod_{i=1}^{N} |\downarrow_{\rm{E}_i}\rangle
\end{equation}
evolves into a $|\psi(t)\rangle$ such that the reduced state of the qubit
has the form of Eq.(\ref{rhot}) but with

\begin{equation}\label{prodO}
O(t)=\prod_{i=1}^N 
\left( 
1-\frac{2b^2_i}{\Omega_i^2}\sin^2\Omega_it 
\right)  
\end{equation}
where $\Omega_i=\sqrt{B^2+b^2_i}$. The early time overlap is
$O(t)=1-2Nb^2t^2$, where $b^2=\sum_i b_i^2/N$. $O(t)$ decays from $1$ on a
timescale of $\tau^{(1)}=1/\sqrt{N}b$. For $b_i^2\ll B^2$ different
factors in Eq.(\ref{prodO}) oscillate with different
frequencies dispersed in the range $B\pm b^2/B$. For $t\gg 2\pi B/b^2$ the
phases of oscillating terms in different factors become effectively random
and the overlap averages to

\begin{equation}
\bar{O} = 1-\frac{Nb^2}{B^2} = 1-\frac1Q \;.
\end{equation}
For $Q=B^2/Nb^2\gg 1$ the system state remains pure; there is only a
negligible partial decoherence. Fig.2 shows $O(t)$ for a sample of $b_i$'s
at two different values of $B$.

\begin{figure}\label{overlap}
\centerline{\epsfxsize=6 cm \epsfbox{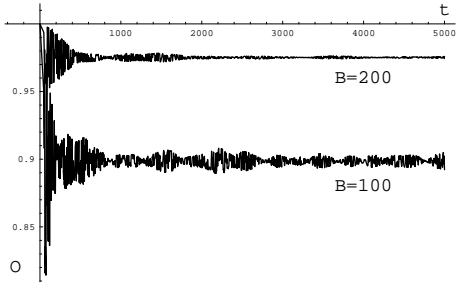}}
\caption{
The overlap $O(t)$ from Eq.(\ref{prodO}) for $N=10^3$ spins with $b^2=1$
at two different values of $B$. $b_i$'s were chosen at random with a
uniform probability distribution between $0$ and $\sqrt{3}$.  The spin
relaxation is not included in this plot or, in other words, $T_r$ is
assumed to be inifinite. 
}
\end{figure}

  Let us include spin relaxation. To get full decoherence it is enough 
that only one out of $N$ spins makes the down transition so the decoherence 
time is $N$ times shorter than for a single spin, 
$\tau^{(2)}=(2T_rB^2/b^2)/N=2QT_r$. $Q\gg 1$ makes this decoherence 
time is much longer than $T_r$.
 
  It is worthwhile to compare the large $Q$ limit with the case of $B=0$
when Eq.(\ref{prodO}) gives

\begin{equation}
O(t)=\prod_{i=1}^N 
\cos 2 b_i t \;. 
\end{equation}
The different factors, which oscillate with frequencies in a range
proportional to $b$, go out of phase after a time of $2\pi/b$ when the
overlap averages to zero. For $B=0$ the time $\tau^{(1)}=1/\sqrt{N}b$ when
$O(t)$ decays away from $1$ is a time of full decoherence.

\section{ Thermalized environment }

  So far we considered the single spin environment and the multispin
environment with a fully polarized initial state. A thermal initial state
is a weighted average over different partially polarized states, like e.g.
$|\uparrow_{\rm{E}_1}\rangle
 |\downarrow_{\rm{E}_2}\rangle
 \dots
 |\uparrow_{\rm{E}_N}\rangle$. The overlap (\ref{prodO}) is the same 
for any such partially polarized state as for the fully polarized state
so the estimates of decoherence time for the polarized state go through 
unchanged for the thermal density matrix.

  The mixedness of the initial environmental state does matter when we
finally take into account the $z$-component of the qubit magnetic field at
the location of spins which gives an extra term in the interaction
Hamiltonian

\begin{equation}
H^z_{\rm{Q-E}}=
\sigma_z \sum_{i=1}^N b_i^z \sigma_z^{\rm{E}_i}. 
\end{equation}
This Hamiltonian, when applied to a partially polarized spins state, gives
$H^z_{\rm{Q-E}}=\sigma_z \sum_{i=1}^N |b_i^z| p_i$, where $p_i=\pm 1$ is a
polarization of the $i$-th spin. If $p_i$'s were static, then this
$H^z_{\rm{Q-E}}$ could be measured at the beginning of the quantum
computation and balanced by a $\sigma_z$-counterterm in the qubit
Hamiltonian, as suggested in Ref.\cite{PCQdec}. The problem is that $p_i$'s
are not static because they flip at the rate of one spin per $T_r/N$. The
total polarization $p(t)=\sum_{i=1}^N p_i(t)$ performs a random walk with
one $\pm 2$ step every $T_r/N$ so that $\overline{[p(t)-p(0)]^2}=4tN/T_r$,
where the overline means here an average over random walks. The
Hamiltonian $H^z_{\rm{Q-E}}\approx\sigma_z b [p(t)-p(0)]$ is random so the
qubit states $|\uparrow\rangle$ and $|\downarrow\rangle$ accumulate
opposite random phases $\pm \phi(t)$, which grow as
$\overline{\phi(t)^2}=4Nb^2t^3/3T_r$. They lead to full dephasing after a
dephasing time $\tau^{(3)}=(3\pi^2T_r/Nb^2)^{1/3}$. In this discussion we
assume $Q\gg 1$ so that entanglement is negligible.

\section{ Numerical estimates }

  The external magnetic field $B$ is bounded from above by the critical
field of the type I superconducting material $H_c$. This parameter is the
highest ($0.206\;T$) for niobium.  In this respect niobium is much better
than alluminium which has much lower $H_c=0.01\;T$. We choose niobium and
set $B=0.1\;T$.  There are materials like lead or tin that have abundant
isotopes with zero nuclear spin. We do not consider them here because
niobium and alluminium microtechnology is more advanced and, what is even
more important, even lead or tin or their substrates would have spin
impurities requiring environment engineering. 
 
  In the qubit design of Ref.\cite{PCQ} the $1\mu m\times 1\mu m$
superconducting loop is made of an alluminium $0.5\mu m\times 0.5 \mu m$
wire with persistent current of $100 nA$. We suggest niobium instead of
alluminium and the loop to be just $0.1 \mu m$ thick. $0.1 \mu m$ is
roughly twice the penetration depth so that $B$ can penetrate through the
superconductor. Note that the critical field is higher in such thin films
than in bulk superconductor \cite{book}. Our superconducting loop will
produce a flux of $10^{-4}\Phi_0$ which is an order of magnitude less than
in the design of Ref.\cite{PCQ}. The loop is also thinner so it contains
fewer spins. The superconductor volume of $10^{-19} m^3$ contains
$10^{10}$ of niobium nuclear spins. They are subject to an average
$b^2=\;10^{-15}T^2$. The quality factor $Q=B^2/Nb^2=10^3$ gives the
overlap $\bar{O}=0.999$. 

  If the external field were $B=0$, then full decoherence would take place
after

\begin{equation}
\tau^{(1)} =
\frac{1}{\sqrt{N}b} \approx 
10^{-5}s \;,
\end{equation}
where we assume the nuclear spin magnetic moment of $10^7 Hz/T$.  For our
$B=0.1T$ after $\tau^{(1)}$ we get only negligible partial decoherence
with $\bar{O}=0.999$. The full decoherence due to spin relaxation could
happen after

\begin{equation}
\tau^{(2)} = 2QT_r \approx 10^5 s\;,
\end{equation}
where we assume $T_r$ to be of the order of minutes, but
the dephasing time  

\begin{equation}
\tau^{(3)} =
\left(\frac{3\pi^2T_r}{Nb^2}\right)^{1/3}
\approx 10^{-3}s, 
\end{equation}
is much shorter. Even this $1ms$ time can be significantly increased by
decreasing temperature because $T_r\sim\exp(\Delta/k_{\rm{B}}T)$, where
$\Delta$ is the superconducting gap. Reduction of temperature from $50mK$
down to $5mK$ increases $\tau^{(3)}$ 25 times.

\section{ Conclusion }

  Maximalization of the quality factor $Q=B^2/Nb^2$ so that $Q\gg 1$ makes
entanglement with spin environment negligible. The same technique can be
used for any qubit design which involves superposition of states with
different magnetic fields. According to our estimates, the present day
persistent current qubit proposal \cite{PCQ} is not far from achieving a
$1ms$ dephasing time due to the finite spin relaxation time $T_r$. $1ms$
is sufficient for the planned quantum computation at a $GHz$ rate. To
achieve this desired decoherence time it is important to replace
alluminium by niobium and make the circuit more microscopic than in
Ref.\cite{PCQ}. It is a common wisdom that more microscopic systems are
better from the point of view of decoherence, compare e.g. our formula for
$\tau^{(1)}$. We showed that in addition to this much expected behaviour,
there is a threshold below which a sufficiently microscopic qubit can be
made free from entanglement with the spin bath.

{\bf Acknowledgements. } I would like to thank Diego Dalvit and Nikolay
Prokof'ev for their comments on the manuscript. This work was supported in
part by NSA.

\end{document}